\documentstyle[11pt]{article}
\newcommand{\beq}{\begin{equation}}
\newcommand{\eeq}{\end{equation}} 

\newcommand{\beqa}{\begin{eqnarray}}
\newcommand{\eeqa}{\end{eqnarray}}
\def\half{\frac{1}{2}}
\def\F{{\cal F}}
\def\opone{\leavevmode\hbox{\small1\kern-3.8pt\normalsize1}}
 
\begin{document}

\title{Quantum cloning without signaling}
\author
{N. Gisin\\
\protect\small\em Group of Applied Physics, University of Geneva, 1211
Geneva 4, Switzerland \\
}
\date{\today}

\maketitle

\begin{abstract}
Perfect Quantum Cloning Machines (QCM) would allow to use quantum nonlocality
for arbitrary fast signaling. However perfect QCM cannot exist. 
We derive a bound on the fidelity of QCM
compatible with the no-signaling constraint. This bound equals the fidelity of
the Bu\v{z}ek-Hillery QCM.
\end{abstract}


Quantum mechanics is non-local, but this cannot be used to signal. Indeed, the different
mixtures of pure states that can be prepared at a distance cannot be distinguished. For
example, if two distant spin $\half$ particle are in the singlet state, then a measurement
of $\vec m\vec\sigma$ on one of them results in a mixture for the second particle spin state:
$\pm\vec m$ on the Bloch sphere, with the same probability $\half$ for both signs. This
mixture corresponds to the density matrix $\half\opone$, independently of the measurement
direction $\vec m$. If one could, in some way or another, distinguish between different
mixtures that can be prepared at a distance, then quantum non locality could be used for
signaling and the peaceful coexistence between quantum mechanics and 
relativity\cite{peacefulCoexistence} would be broken. If, for example,
perfect quantum cloning machines (QCM) would exists, then, by cloning the second particle, 
the mixtures corresponding to different directions $\vec m$ could be distinguished. But
perfect QCM do not exist\cite{noCloningTheorem}. 
However, imperfect QCM exist\cite{BuzekQCM,GisinMassarQCM,BrussQCM}. 
In this letter we derive the upper
bound of the quality (fidelity) of QCM of qubits (spin $\half$) compatible with no
signaling: any hypothetical QCM with higher fidelity would allows signaling.

Let $\rho_{in}(\vec m)=\frac{1+\vec m\vec\sigma}{2}$ and $\rho_{out}(\vec m)$ denote 
the input and the corresponding output state of our 
hypothetical QCM, with $\rho_{out}(\vec m)$ a 2-qubit
state. We assume that the 2 output qubits are in the same state: $Tr_1(\rho_{out}(\vec m))=
Tr_2(\rho_{out}(\vec m))=\frac{1+\eta\vec m\vec\sigma}{2}$, where $\eta$ is the shrinking
factor of the Bloch vector $\vec m$. The cloning fidelity is given by: 
$\F\equiv Tr(\rho_{in}(\vec m)\frac{1+\eta\vec m\vec\sigma}{2})=\frac{1+\eta}{2}$.
Let us emphasize that we do not assume any further particular relation between
$\rho_{in}(\vec m)$ and $\rho_{out}(\vec m)$, in particular we do not assume any linear 
relation.
In full generality, the output state can be written as:
\beq
\rho_{out}(\vec m)=\frac{1}{4}(\opone+\eta(\vec m\vec\sigma\otimes\opone+\opone\otimes\vec m\vec\sigma)
+\sum_{j,k=x,y,z}t_{jk}\sigma_j\otimes\sigma_k)
\label{rhoout1}
\eeq

By definition, universal QCM act similarly on all input states:
\beq
\rho_{out}(U\vec m)=U\otimes U\rho_{out}(\vec m)U^\dagger\otimes U^\dagger
\eeq
for all $U$, where the same notation $U$ is used for unitary operators acting on the
2-dimensional spin $\half$ Hilbert space and for the corresponding rotation operator acting
on the Bloch vectors $\vec m$. A first consequence of this covariance property is that
$\rho_{out}(\vec m)$ is invariant under rotation around the direction
$\vec m$: $[e^{i\alpha\vec m\vec\sigma}\otimes e^{i\alpha\vec m\vec\sigma}, \rho_{out}(\vec m)]=0$
for all $\alpha$.
This imposes conditions on the $t_{jk}$ parameters. For example, if $\vec m$ is in the z-direction,
then these conditions read: $t_{xx}=t_{yy}$, $t_{xy}=-t_{yx}$ and $t_{xz}=t_{zx}=t_{yz}=t_{zy}
=0$. Thus:             
\beqa
\rho_{out}(\uparrow)&=&\frac{1}{4}(\opone+\eta(\sigma_z\otimes\opone+\opone\otimes\sigma_z)  \\ \nonumber
&+&t_{zz}\sigma_z\otimes\sigma_z + t_{xx}(\sigma_x\otimes\sigma_x+\sigma_y\otimes\sigma_y)
+t_{xy}(\sigma_x\otimes\sigma_y-\sigma_y\otimes\sigma_x))  \\
&=&\frac{1}{4}\pmatrix{1+2\eta +t_{zz} & 0 & 0 & 0\cr 0 & 1-t_{zz} & 2t_{xx}+2it_{xy} & 0\cr 
0 & 2t_{xx}-2it_{xy} & 1-t_{zz} & 0\cr 0 & 0 & 0 & 1-2\eta+t_{zz}}
\label{rhoout2}
\eeqa

Now, the no signaling condition implies \cite{noSignalcond}:
\beq
\rho_{out}(\uparrow)+\rho_{out}(\downarrow)=\rho_{out}(\rightarrow)+\rho_{out}(\leftarrow)
\label{nosignal}
\eeq
This implies that $t_{zz}=t_{xx}=t_{yy}\equiv t$.

Finally, the last condition that $\rho_{out}$ has to fulfil is positivity. The eigenvalues of
$\rho_{out}(\uparrow)$ are:
\beqa
\frac{1}{4}(1\pm2\eta +t)  \\
\frac{1}{4}(1-t\pm2\sqrt{t^2+t_{xy}^2})
\eeqa
The maximum value of $\eta$ under the condition that these 4 eigenvalues are non-negative is
obtained for $t_{xy}=0$ and $t=\frac{1}{3}$. Hence $\eta_{max}=\frac{2}{3}$ and
the maximum fidelity is given by:
\beq
\F_{max}=\frac{1+\eta_{max}}{2}=\frac{5}{6}
\eeq
This fidelity is obtained by the Bu\v{z}ek-Hillery QCM\cite{BuzekQCM}. 
This proves that the  Bu\v{z}ek-Hillery QCM is optimal, as already shown in 
\cite{GisinMassarQCM,BrussQCM,WernerQCM}.

A quite straightforward proof of the optimality of the Bu\v{z}ek-Hillery QCM\cite{BuzekQCM}
has been presented, based on the fact that no quantum process can provide arbitrary fast
signaling\cite{noSignaling}.
Once again, quantum mechanics is
right at the border line of contradicting relativity, but does not cross it. The peaceful
coexistence between quantum mechanics and relativity\cite{peacefulCoexistence} 
is thus re-enforced. It is intriguing that the no signaling constraint is a powerful
guide to find the limits of quantum mechanics\cite{GisinRigoRelevant}.

\section*{Acknowledgments}
Support by the Swiss National Science Foundation is gratefully acknowledged. Numerous
discussions with Bruno Huttner, Helle Bechmann and many participants to the European
TMR network "The Physics of Quantum Information" have stimulated this work. I also
profited from the stimulating ISI workshops sponsored by Elsag-Bailey.


\begin{thebibliography}{99}
\bibitem{noSignalcond} The no signaling condition imposes that the mixtures of output
       states corresponding to indistinguishable mixtures of input states are themselves
       indistinguishable. This is more general than the condition used here. However, it
       turns out that in the present context, the simple condition (\ref{nosignal}) is sufficient.
\end{thebibliography}
\end{document}